\documentstyle[eqsecnum,pra,aps]{revtex}
\begin{document} \draft
\title {Temporally stable coherent states in energy degenerate systems:
The hydrogen atom}
\author{M.G.A.\ Crawford\cite{byline}}
\address{Department of Applied Mathematics, University of Waterloo,\\
Waterloo, Ontario, Canada, N2L 3G1}
\maketitle
\begin{abstract}
Klauder's recent generalization of the harmonic oscillator coherent
states [J. Phys. A 29, L293 (1996)] is applicable only in non-degenerate
systems, requiring some additional structure if applied to systems
with degeneracies.  The author suggests how this structure could be
added, and applies the complete method to the hydrogen atom problem.
To illustrate how a certain degree of freedom in the construction may
be exercised, states are constructed which are initially localized
and evolve semi-classically, and whose long time evolution exhibits
``fractional revivals.''
\end{abstract}


\widetext

\section{Introduction}

Since their early introduction to quantum mechanics, the harmonic
oscillator coherent states \cite{klauder85} have served many
purposes.  In Schr\"odinger's conception \cite{schrodinger26}, they
were viewed as quasi-classical objects, purely quantum in definition
though remarkably classical in behaviour.  From this perspective,
some authors have used these states with a classical limit to study
correspondence between quantum and classical perturbation series
\cite{mcrae97}.  Stemming from Glauber's study \cite{glauber63c},
coherent states have a wide application in quantum optics
\cite{icssur99}, primarily in representations of the electromagnetic
field.  In fact, being continuously parameterized, coherent states
figure prominently in the theory of continuous representations
\cite{klauder63a,klauder63b}.

Many generalizations of coherent states appear in the literature.
Each generalization tends to preserve a small number of the properties
of the harmonic oscillator coherent states in the general scheme
at the expense of the remaining properties.  A recent generalization
due to Klauder \cite{klauder96} preserves many, at the expense of
few.  Klauder's generalization gives states which (a) evolve among
themselves in time (temporally stable), (b) are continuously
parameterized, and (c) admit a resolution of the identity.  As
such, no reservations are made for ``semi-classical'' properties
such as minimum uncertainty, though a certain degree of freedom to
be discussed below remains within the construction which may be
optimized according to additional concerns.  Two studies have since
appeared \cite{majumdar97,klauder99} proposing fourth conditions
which eliminate this degree of freedom to be discussed in Section
\ref{clair}.

As initially presented, Klauder's construction is appropriate for
systems without energy degeneracies.  With no additional
structure, the resolution of the identity fails for degenerate
systems.  Energy degeneracies arise when independent operators
commute with the Hamiltonian, suggesting a Lie algebraic approach
to impose the additional structure.  Thus in the presence of
degeneracies, excepting those few cases of truly ``accidental''
degeneracies, the Perelomov approach to constructing coherent states
for the degeneracy group is an obvious and general path to take
\cite{perelomov86}.

The attempt to generalize harmonic oscillator coherent states for
the hydrogen atom problem is not new.  Many involve the construction
of the complete Perelomov states via the dynamical group SO(4,2)
\cite{mostowski76,gerry86,mcanally90,zlatev94}.  Others, including
the current approach, make some use of SO(4) coherent states
\cite{gay89,nauenberg89}.  Others involve the construction of
``temporally stable'' coherent states:  Klauder's original paper
\cite{klauder96}, a study by Majumdar and Sharatchandra \cite{majumdar97},
and another by Fox \cite{fox99}.  The current construction differs
from these in some key respects which will be pointed out in the
course of the paper.

Section \ref{const} discusses the construction due to Klauder, and
develops the extension of this construction to degenerate states.
In Section \ref{hydrogen}, the construction is applied to the
hydrogen atom problem.  Section \ref{clair} is a discussion regarding
role that these states and other generalized coherent states hold
in physical theory, namely, as representations rather than as
physical states.  The dynamics of an individual state are explored
in Section \ref{dynamics}, in which hydrogenic states are constructed
which exhibit fractional and full revivals, and the main points of
the paper are summarized in Section \ref{concl}.

\section{The Construction}
\label{const}

The generalization due to Klauder \cite{klauder96} is applicable to the
discrete portion of the spectrum of a Hamiltonian $\hat{H}$.  (A
continuum generalization has also appeared \cite{klauder99}.)
So, for a non-degenerate $\hat{H}$ with eigenstates $|n\rangle$ and
eigenstate energies $e_{n}$, the coherent states are given by
(using atomic units)
\begin{equation}
|{s,\gamma}\rangle = M(s^{2}) \sum_{n=0}^{\infty} \frac{s^{n}
\exp(-i \gamma e_{n})}{\sqrt{\rho_{n}}} |{n}\rangle,
\label{kssta}
\end{equation}
where $s \geq 0$ and $\gamma$ is real.  The factors $\rho_{n}$ are the
moments of a function $\rho(u)>0$, $u \geq 0$,
\begin{equation}
\rho_{n} = \int_{0}^{\infty} u^{n} \rho(u) du,
\end{equation}
and the normalizing function $M(s^{2})$ is chosen such that
\begin{equation}
1 = \langle{s,\gamma|s,\gamma}\rangle = M^{2}(s^{2}) \sum_{n=0}^{\infty}
\frac{s^{2n}}{\rho_{n}}.
\label{rcond}
\end{equation}
If the discrete portion of the spectrum is finite, the upper limits of
the above sums (and the appropriate expressions hereafter) may be
replaced with $n_{\mbox{\scriptsize max}}$.

The choice of $\rho(u)$ is the remaining degree of freedom up to the
following restrictions:  All the moments $\rho_{n}$ exist (up to
$n_{\mbox{\scriptsize max}}$ if finite), and the sum in Eq.\ (\ref{rcond})
exists for all $s \geq 0$.  Such functions $\rho(u)$ are known to exist:
In the harmonic oscillator, $\rho(u)=e^{-u}$ leads to the
standard harmonic oscillator coherent states.

The states $|s,\gamma\rangle$ clearly are 
continuously parameterized.  The states exhibit temporal
stability: $\exp(-i \hat{H} t)|s,\gamma\rangle = |s,\gamma+t\rangle$.
Further, given $\rho(u)$ and $M^{2}(u)$, let $k(u)$ be defined by
\begin{equation}
k(u) M^{2}(u) = \rho(u).
\end{equation}
Then, the coherent states satisfy the resolution of the identity,
\begin{equation}
\hat{1} = \int d \mu(s,\gamma) |{s, \gamma}\rangle \langle{s,\gamma}|,
\label{id}
\end{equation}
in which the integration is given by
\begin{equation}
\int d \mu(s,\gamma) = \lim_{\Gamma \rightarrow \infty} \frac{1}{2 \Gamma}
\int_{0}^{\infty} ds^{2} k(s^{2}) \int_{-\Gamma}^{\Gamma} d\gamma.
\end{equation}
The limit is necessary to accommodate possible incommensurabilities
of energy levels.  In proving Eq.\ (\ref{id}), integration
over $\gamma$ yields the Kronecker delta $\delta_{e_{n}e_{m}}$,
which may be identified with $\delta_{nm}$ only in the non-degenerate
case.  Also, the identity of Eq.\ (\ref{id}) should be regarded as
a projection operator onto the states contributing to $|{s,\gamma}\rangle$,
{\em i.e.} the discrete portion of the spectrum.  These properties in
concert make these states most useful in the representation of
arbitrary, bound, time evolved states.

To extend the construction to energy degenerate states, replace
Eq.\ (\ref{kssta}) with
\begin{eqnarray}
\lefteqn{ |{s,\gamma,x}\rangle} \nonumber \\
&=& N(s^{2}) \sum_{n=0}^{\infty} \frac{s^{n} \exp(-i \gamma
e_{n} )}{\sqrt{\rho_{n}}} \sqrt{d_{n}} |{n,x}\rangle,
\label{gencs}
\end{eqnarray}
where $d_{n}$ is the degeneracy of the $n$th energy level,
$|{n,x}\rangle$ are the Perelomov coherent states \cite{perelomov86}
for the degeneracy group $G$, and the normalizing factor $N(s^{2})$
is given by
\begin{equation}
1 = \langle{s,\gamma,x|s,\gamma,x}\rangle = N^{2}(s^{2}) \sum_{n=0}^{\infty}
\frac{s^{2n} d_{n}}{\rho_{n}} .
\end{equation}
In each energy degenerate subspace of the Hilbert space,
the Perelomov coherent states satisfy the resolution of the identity
\begin{equation}
\hat{1}_{n} = d_{n} \mbox{vol}(H) \int_{X} d \eta(x)
|{n,x}\rangle \langle{n,x}|,
\label{dada}
\end{equation}
in which $H$ is the isotropy subgroup relative to the fiducial
vector in the construction of the Perelomov coherent states, $X = G/H$
is the quotient space formed by the degeneracy group with the
isotropy subgroup, and the measure $d \eta$ is induced from the
Haar measure on the degeneracy group.  The states Eq.\ (\ref{gencs})
therefore satisfy the resolution of the identity
\begin{equation}
\hat{1} = \int d\mu(s,\gamma,x) |s,\gamma,x\rangle\langle s,\gamma,x|,
\end{equation}
with
\begin{eqnarray}
\lefteqn{\int d\mu(s,\gamma,x)  = } \\
& & \lim_{\Gamma \rightarrow \infty} \frac{1}{2 \Gamma}
\int_{0}^{\infty} ds^{2} k(s^{2}) \int_{-\Gamma}^{\Gamma} d\gamma \;
\mbox{vol}(H)\int_{X} d\eta(x).
\nonumber
\end{eqnarray}

Since the states $|{n,x}\rangle$ are formed by superpositions over
states which share a common energy eigenvalue $e_{n}$, they are
also eigenstates of the Hamiltonian and so evolve simply in
time.  Accordingly, the states $|{s,\gamma,x}\rangle$ preserve the
temporal stability property of the non-degenerate construction.

Majumdar and Sharatchandra's construction \cite{majumdar97} also
makes explicit use of the Perelomov construction of coherent states
for the degeneracy group, and Klauder's construction \cite{klauder96}
is less explicit in this regard.  A significant difference between
this and these other extensions of Klauder's construction is the
treatment of the factor $d_{n}$.  Among other problems \cite{bellomo99}
Majumdar and Sharatchandra incorporate $d_{n}$ into the measure
after the summation of the state, an operation of questionable
justifiability.  Klauder, using an adaptation of SO(3) coherent
states, incorporates $d_{n}$ into the states after, and therefore
affecting, normalization.  Fox \cite{fox99} constructs temporally
stable coherent states, but with a Gaussian taking the place
of $s^{n} \exp (-i \gamma e_{n})$ in Eq.\ (\ref{kssta}).  As such,
a similar adaptation of his states may be effected by including
a factor of $\sqrt{d_{n}}$ as in Eq.\ (\ref{gencs}).

\section{Special case: The hydrogen atom}
\label{hydrogen}

The group theoretical treatment of the hydrogen atom is standard
in the literature \cite{barut71a,cizek82,adams88}.  For the
hydrogen atom problem, there are two realizations of
the degeneracy group SO(4).  One uses the elements of the angular
momentum vector, $\hat{L}_{j}$, and a scaled quantum Runge-Lenz
vector, $\hat{A}_{j}$, as generators of the group, whereas the
other decouples these six generators into two sets, $\hat{M}_{j}
= \frac{1}{2} (\hat{L}_{j}+\hat{A}_{j})$, and $\hat{N}_{j} =
\frac{1}{2} (\hat{L}_{j}-\hat{A}_{j})$.  In the second representation,
one finds that $\mbox{SO}(4) = \mbox{SO}(3) \otimes \mbox{SO}(3)$,
so that, loosely speaking, a Perelomov coherent state for SO(4) may be
given by the direct product of two SO(3) coherent states.

The SO(3) coherent states with the fiducial vector $|j,-j\rangle$,
$j=0,\frac{1}{2},1,\frac{3}{2},\ldots$, are given by
\begin{equation}
|{j,\zeta}\rangle =
\sum_{m=-j}^{j} \left [ \frac{(2j)!}{(j+m)!(j-m)!} \right ]^{1/2}
\frac{\zeta^{j+m}} {(1+|\zeta|^{2})^{j}} |{j,m}\rangle.
\end{equation}
The resolution of the identity for these states may be written
\begin{equation}
\hat{1}_{j} = \frac{2j+1}{\pi} \int \frac{d^{2} \zeta}{(1+|\zeta|^{2})^{2}}
|{j,\zeta}\rangle \langle{j,\zeta}|,
\label{zorks}
\end{equation}
where $d^{2} \zeta = d \Re \zeta d \Im \zeta$, and integration is over
the entire complex plane projected from the unit sphere with the
transformation $\zeta = - \tan \frac{\theta}{2} e^{-i \phi}$.
In the hydrogenic realization, the representations of each copy of
SO(3) are of equal dimension ($n=2j+1$) so the dimensions of the
relevant representations of SO(4) are $n^2$, $n=1,2,3,\ldots$.

It is now straight-forward to construct the coherent states for the
full system.  The coherent states for the hydrogen atom problem by
this construction are given by
\begin{eqnarray}
\lefteqn{|{s,\gamma,\zeta_{1},\zeta_{2}}\rangle =} \\
& & N(s^{2}) \sum_{n=0}^{\infty} \frac{s^{n} \exp (-i \gamma e_{n+1}) (n+1)}
{\sqrt{\rho_{n}}} |{n+1,\zeta_{1},\zeta_{2}}\rangle
\nonumber
\end{eqnarray}
with,
\widetext
\begin{equation}
|n,\zeta_{1},\zeta_{2}\rangle = \sum_{m_{1},m_{2}=-j}^{j}
\frac{(2j)!\zeta_{1}^{j+m_{1}}\zeta_{2}^{j+m_{2}}
|{j,m_{1}}\rangle|{j,m_{2}}\rangle}
{[(j+m_{1})!(j-m_{1})!(j+m_{2})!(j-m_{2})!]^{1/2}
(1+|\zeta_{1}|^{2})^{j} (1+|\zeta_{2}|^{2})^{j}} .
\label{fif}
\end{equation}
The states $|{j,m_{1}}\rangle|{j,m_{2}}\rangle$ may be related to
the standard hydrogen Hamiltonian eigenstates $|n,\ell,m\rangle$
via Clebsch-Gordon coefficients.  The states Eq.\ (\ref{fif})
satisfy the resolution of the identity
\begin{equation}
\hat{1}_{B} = \frac{1}{\pi^{2}} \int d\mu(s,\gamma) \int \frac
{d^{2}\zeta_{1} d^{2}\zeta_{2}}
{(1 + |\zeta_{1}|^{2})^{2} (1 + |\zeta_{2}|^{2})^{2}}
|{s,\gamma,\zeta_{1},\zeta_{2}}\rangle
\langle{s,\gamma,\zeta_{1},\zeta_{2}}|,
\label{hydres}
\end{equation}
where the subscripted $B$ is included to emphasize that this is
more appropriately regarded as a projection operator into the bound
portion of the Hilbert space.  In the specific example of $\rho(u)
= e^{-u}$, with moments $\rho_{n} = n!$, explicit form may be given
to $N(s^{2})$ and $k(u)$ by
\begin{equation}
N(s^{2}) = e^{-s^{2}/2} (1 + 3 s^{2} + s^{4} ) ^{-1/2}
\end{equation}
and
\begin{equation}
k(u) = 1 + 3 u + u^{2}.
\end{equation}

\section{Some Clarification}
\label{clair}

At this point, a few observations are in order.  Primarily, the
term ``temporal stability'' in no way refers to the time evolution
of the structure in configuration space.  Only through a rather
generous interpretation does this construction ``positively'' solve
the long standing problem of forming non-dispersing wave packets
for the hydrogen atom.  Temporal stability refers strictly to the
mathematical property that the states evolve in time among themselves.
With this property in mind, some authors \cite{majumdar97} have
grossly overstated the the nature of the configuration space time
evolution, while other authors \cite{bellomo98,bellomo99} have
studied in detail the long-time evolution of individual states,
even though there is no underlying physical basis either to provide
for spatial coherence, or to presume states of this description
are found in the laboratory at all.  The question of how to prepare
these states in the laboratory remains very much open.

Much of the study of generalized coherent states rests more in the
mathematical than the physical nature of mathematical physics.
Glauber's motivation in his study of coherent states \cite{glauber63c}
was not so much that coherent states are found in the laboratory,
but that they provide a representation in which otherwise difficult
calculations become feasible.  Glauber noted that certain electric
field operators have representations as sums over the modal
annihilation operators.  In diagonalizing these operators, one
arrives at eigenstates of the modal annihilation
operators.  Restricted to a single mode, this corresponds to the
annihilation operator definition of harmonic oscillator coherent
states.  Hence as annihilation operator coherent states, they arise
from a representation, a point of mathematical convenience, not as
a conclusion from the physics of the problem.  In any case,
generalizations of annihilation operator coherent states have
appeared widely, though the physical motivation to study such
definitions in any context besides as representations is unclear.

Glauber also showed how these states may be constructed through
the action of a displacement operator on the ground state.  This
definition was generalized by Perelomov \cite{perelomov72}, a
generalization which has been widely successful.  This success is
founded upon the properties of the dynamical group coming through
into the set of coherent states, not from an assertion (which few
researchers make) that an individual state by such a construction
matches a state by some preparation in the laboratory.  This success
is of a mathematical, not physical, nature again resting upon the
use of these states as a representation.

Of Glauber's original three definitions, the approach which appears
to invest the most physics is the minimum uncertainty construction.
Indeed, squeezed states, a generalization of this construction,
are used as descriptions of physical aspects of certain quantum
optical experiments in the laboratory.  Nieto {\em et al.}
\cite{nieto81} have also developed a generalization which minimizes
the uncertainty product of a pair of ``natural'' operators.
Ehrenfest's relations then lead to the initial evolution of the
quantum expectation values approximating classical evolution.
Though this approach is strongest in terms of an underlying physical
motivation, these states still lack (in general) any physical
hypothesis which selects for states of this description in the
laboratory.  As an aside, the Nieto construction, though seen from
time to time, is not as widely used as the Perelomov construction
for perhaps two reasons.  Firstly, though it is generally applicable
in principle, many systems are intractable to carry through to
completion (when the Hamiltonian enters into the ``natural''
operators).  Secondly, a certain tradeoff appears to be at work:
This approach is somewhat less mathematically endowed than
Perelomov's approach.

Now consider Klauder's construction.  All of the attractions are
mathematical in nature.  As initially presented, no reservations
are made for coherence in configuration space ({\em i.e.} semi-classical
behaviour) and there is no general physical mechanism which would
result in finding these states in the laboratory.  However, a
certain degree of freedom remains in the construction, and two
suggestions have separately appeared that a fourth requirement will
simultaneously eliminate the degree of freedom and ensure for the
behaviour in configuration space.  It is likely that a fourth
requirement, if it exists, will be physical in nature.  The
requirement postulated by Majumdar and Sharatchandra \cite{majumdar97}
is that the measure found in the resolution of the identity
corresponds to the ``canonical'' measure on classical phase space.
They further assert that the measure uniquely identifies the set
of coherent states.  This assertion is false, as shown by Sixdeniers
{\em et al.} \cite{sixdeniers99} who demonstrate multiple measures
corresponding to the same set of coherent states. Also, though it
is convenient from a mathematical point of view, it is unclear why
the measures should correspond at all from a physical point of
view, or even if a meaningful identification (one to one) can always
be made between individual coherent states and points in classical
phase space.

A fourth requirement is also postulated by Gazeau and Klauder
\cite{klauder99} which is motivated by an attempt to formalize the
connection between the quantum parameters to the coherent state
and the classical action-angle variables.  Unfortunately, their
requirement results in an angle variable whose rate of change with
time is independent of the action, a rather special circumstance
restricted to the harmonic oscillator and a small number of other
systems.  This is a severe limitation in terms of semi-classical
behaviour, since this is clearly at odds with how the angle variable
evolves in, say, the hydrogen atom problem.

A degree of freedom also remains in Fox's construction \cite{fox99}
of Gaussian generalized coherent states, namely the width
of the Gaussian in question.  Note that in this context, the
distribution in energy level, not configuration space, is Gaussian.
Fox does not give any criteria which are intended to specify a
suitable width.  As with the Klauder's construction, this degree
of freedom may be optimized according to the aim in mind.

Hence in the absence of an acceptable fourth criterion (none is
herein proposed), we carry on.  This limits the construction to a
mathematical tool, though an interesting mathematical tool it is.
Note that the time dependent Schr\"odinger equation for a
time-independent Hamiltonian in this coherent state representation
becomes (in atomic units)
\begin{equation}
\frac{\partial}{\partial \gamma} \langle s,\gamma | \psi \rangle =
- \frac{\partial}{\partial t} \langle s,\gamma | \psi \rangle
\end{equation}
whose solution is trivial.  With this expression representing the
unperturbed solution, this would make for an interesting point of
departure towards a study of time dependent perturbation theory.
That is, coherent states are useful when considered as an ensemble
of states (a representation), not as individual states.

Speaking now in the specific case, some authors \cite{bellomo98,bellomo99}
have suggested that the temporally stable construction of coherent
states does not support the possibility of exhibiting full or
fractional revivals as described by Averbukh and Perelman
\cite{averbukh89} or Nauenberg \cite{nauenberg89}.  Firstly, in order to
decide whether a state is to be found in the laboratory, a physical
mechanism for the preparation of these states must be postulated.
Until this has been done, the presence or absence of a phenomenon which is,
after all, universal is not relevant.  Secondly, these authors did
not exploit the degree of freedom which remains in the construction.
Without supplying a physical motivation which would lead to finding
these states in the laboratory, we shall see that by exploiting
this degree of freedom wave functions may be formed by the present
construction which exhibit the full panoply of revivals.

\section{Dynamics}
\label{dynamics}

Having thus constructed the states emphasizing, among other things,
time evolution, it is now interesting to consider the behaviour of
the states as evolved in time.  Other authors have defined hydrogen
atom coherent states with a variety of constructions and with
various reports of evolution in ``fictitious'' time \cite{gerry86},
or evolution along circular \cite{brown73,gaeta90} or Keplerian
elliptical orbits with possible, eventual state revivals
\cite{gay89,nauenberg89}.  Coherent states also may be constructed
by the present recipe which travel along elliptical orbits and
exhibit fractional revivals.

According to Averbukh and Perelman \cite{averbukh89}, fractional
revivals are a universal phenomenon exhibited by wave functions
provided third order corrections and higher do not contribute
significantly to a polynomial approximation to the energy eigenvalues
over contributing energy eigenstates.
Expanding about $n=\bar{n}$, the hydrogen atom energy levels are 
\begin{eqnarray}
\lefteqn{ e_{n} = - \frac{1}{2 n^{2}} = } \\
& & \mbox{} - \frac{1}{2 \bar{n}^{2}} + \frac{1}{\bar{n}^{3}}(n-\bar{n}) -
\frac{3}{2 \bar{n}^{4}}(n-\bar{n})^{2}
+ \frac{2}{\bar{n}^{5}}(n-\bar{n})^{3} + \cdots.
\nonumber
\end{eqnarray}
The first two terms lead to phase angles equal to multiples of $2\pi$
for times in the vicinity of
\begin{equation}
t=T_{r}=\frac{2\pi}{3} \langle n \rangle ^{4}.
\label{tr}
\end{equation}
This signals a wave function revival at $t=T_{r}$ provided the cubic
term is small at the edges of the distribution, {\em i.e.,}
$4\pi(\Delta n)^{3} \ll 3 \langle n \rangle$.
Note that $T_{r}=T_{rev}/2$ in Averbukh and Perelman's notation.

At the heart of these coherent states
lies the function $\rho(u)$.  With $\rho(u)=e^{-u}$,
one obtains a distribution in eigenlevels characterized by
\begin{equation}
\langle n \rangle = s^2 \frac{s^{4}+5s^{2}+4}{s^{4}+3s^{2}+1}
\end{equation}
and
\begin{equation}
(\Delta n)^{2} = s^2 \frac{s^{8}+6s^{6}+14s^{4}+10s^{2}+4}
{s^{8}+6s^{6}+11s^{4}+6s^{2}+1},
\end{equation}
so that, taking leading order behaviour, $\Delta n \sim \sqrt{\langle
n \rangle}$.  Substituting into the above necessary condition for a
revival at $t=T_{r}$ gives $4 \pi \sqrt{\langle n \rangle} / 3 \ll 1$
which is only satisfied for states in the immediate vicinity of the
trivial coherent state, the ground state, and certainly violated by
states with high principle quantum numbers.

However, the function $\rho(u)$ is a ``degree of freedom'' in the
construction, and may be chosen according to application in mind.
Accordingly, consider instead $\rho(u)=\exp(-u^{\alpha})$ for some
constant $\alpha > 0$ with a view to constructing wave packets
which exhibit strong revivals.  The moments of this function are
\begin{equation}
\rho_{n} = \frac{1}{\alpha} \Gamma (\frac{n+1}{\alpha}).
\end{equation}
Many of the expressions involved in subsequent calculations may be
handled using properties of the functions of Mittag-Leffler
\cite{bateman81}, though they will be treated instead by comparisons
to expressions following from $\rho(u)=e^{-u}$.  In fact, using
$\rho(u)=\exp(-u^{\alpha})$ results in a set of coherent states
closely related to those described by Sixdeniers {\em et al.}
\cite{sixdeniers99}.

Expressions for $\langle n \rangle$ and $(\Delta n)^{2}$ may be
approximated by recognizing the scalings necessary to map expressions
with $\alpha=1$ onto those with general $\alpha$:
\begin{equation}
n+1  \rightarrow  (n+1)/\alpha, \quad
s  \rightarrow  s^{\alpha}.
\end{equation}
Hence, one obtains to leading order (large values of $s$ will
eventually be involved)
\begin{equation}
\langle n \rangle \sim \alpha s^{2 \alpha} , \quad 
\Delta n  \sim \alpha s^{\alpha},
\end{equation}
so that $\Delta n \sim \sqrt{\alpha \langle n \rangle}$.  Substituting
this into the minimal condition for the first full revival gives
$4 \pi \alpha^{3/2} \sqrt{\langle n \rangle} / 3 \ll 1$, which may
be satisfied if $\alpha$ is chosen sufficiently small.  Without
discussing the effect of changing $\alpha$ any further, there will
be a tradeoff between large and small $\alpha$:  Large $\alpha$
will introduce many significantly contributing energy levels for
a given $\langle n \rangle$ yielding good spatial localization but
weak or non-existent revivals, whereas small $\alpha$ yields strong
revivals of poorly localized states.  Note that in this construction
a small width in $n$ follows from an appropriate choice for $\rho(u)$
whereas the same may be accomplished by Fox's construction \cite{fox99}
by simply specifying the width to be narrow.

\begin{figure}
\begin{center}
\begin{picture}(0,0)%
\includegraphics{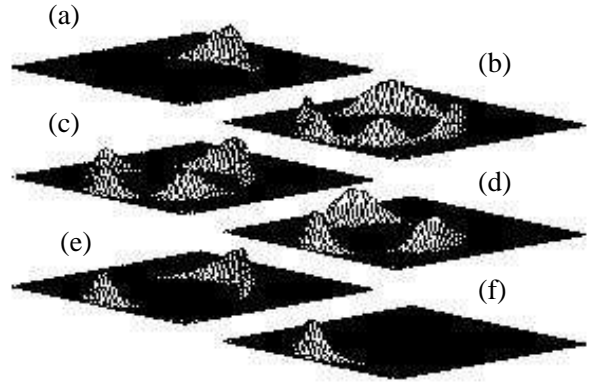}%
\end{picture}%
\setlength{\unitlength}{3947sp}%
\begingroup\makeatletter\ifx\SetFigFont\undefined%
\gdef\SetFigFont#1#2#3#4#5{%
  \reset@font\fontsize{#1}{#2pt}%
  \fontfamily{#3}\fontseries{#4}\fontshape{#5}%
  \selectfont}%
\fi\endgroup%
\begin{picture}(4176,2692)(4639,-2816)
\end{picture}
\caption{Amplitude of coherent state on the $x$-$y$ plane for times
(a) $t=0$, (b) $t=T_{r}/5$, (c) $t=T_{r}/4$, (d) $t=T_{r}/3$, (e)
$t=T_{r}/2$ and (f) $t=T_{r}$.}
\label{peaks}
\end{center}
\end{figure}

As a typical example, consider the state depicted in Figure \ref{peaks}.
For this state, $\alpha = 1/32$ and $s = 2.209 \times 10^{59}$.
This results in a state centred at $\langle n \rangle = 160$ with
a width of $\Delta n = \sqrt{5}$ for which one expects a full
revival at $t=T_{r} = 1.373 \times 10^{9}$.  The parameters
$\zeta_{1,2}$ to the SO(4) coherent state were chosen provide an
elliptical orbit with eccentricity $\epsilon = 0.385$, the major
axis parallel to the $x$-axis, and the angular momentum parallel
to the $z$-axis.  The vertical axes are amplitudes of the wave
functions on the $x$-$y$ plane, calculated at the times indicated
on a square grid 80000 units in width centred at the origin.

The evolution of this state is as expected.  Initially localized,
the state evolves semi-classically.  The wave function then spreads
out but remains close to the ellipse.  As the expected times for
the various fractional revivals arrive, the state exhibits the
expected revival including the full revival at $t=T_{r}$, even though
an examination of the minimal condition for the first revival gives
the debatable result $0.29 \ll 1$.

\begin{figure}
\begin{center}
\begin{picture}(0,0)%
\includegraphics{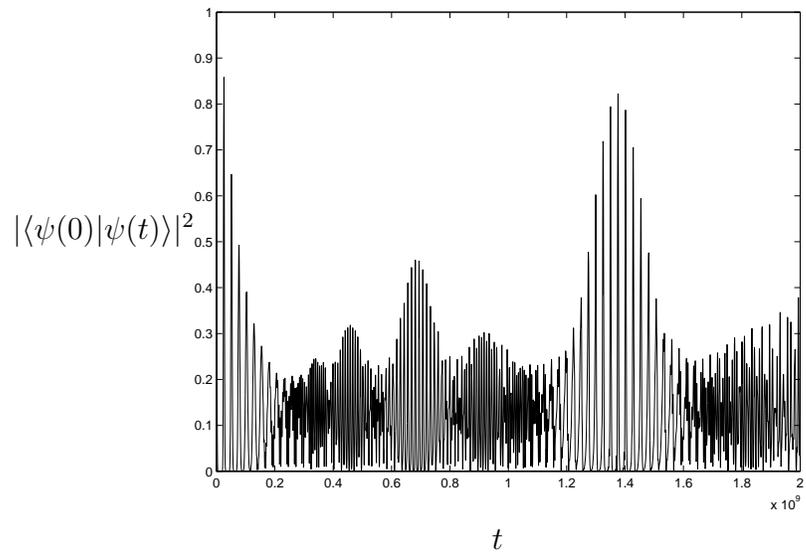}%
\end{picture}%
\setlength{\unitlength}{3947sp}%
\begingroup\makeatletter\ifx\SetFigFont\undefined%
\gdef\SetFigFont#1#2#3#4#5{%
  \reset@font\fontsize{#1}{#2pt}%
  \fontfamily{#3}\fontseries{#4}\fontshape{#5}%
  \selectfont}%
\fi\endgroup%
\begin{picture}(5875,3462)(-762,-4111)
\put(3151,-4111){\makebox(0,0)[b]{\smash{\SetFigFont{12}{14.4}{\familydefault}{\mddefault}{\updefault}$t$}}}
\put(676,-2161){\makebox(0,0)[b]{\smash{\SetFigFont{12}{14.4}{\familydefault}{\mddefault}{\updefault}$|\langle \psi (0)|\psi(t) \rangle|^{2}$}}}
\end{picture}
\caption{Autocorrelation function of the coherent state at time $t$.
For this state, the revival time is $T_{r} = 1.373 \times 10^{9}$.}
\label{auto}
\end{center}
\end{figure}

\begin{figure}
\begin{center}
\begin{picture}(0,0)%
\includegraphics{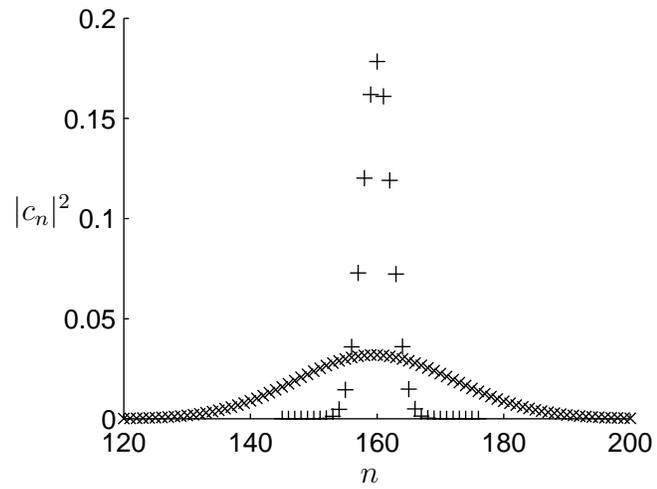}%
\end{picture}%
\setlength{\unitlength}{3947sp}%
\begingroup\makeatletter\ifx\SetFigFont\undefined%
\gdef\SetFigFont#1#2#3#4#5{%
  \reset@font\fontsize{#1}{#2pt}%
  \fontfamily{#3}\fontseries{#4}\fontshape{#5}%
  \selectfont}%
\fi\endgroup%
\begin{picture}(4870,3087)(243,-3361)
\put(1351,-1711){\makebox(0,0)[rb]{\smash{\SetFigFont{12}{14.4}{\familydefault}{\mddefault}{\updefault}$|c_{n}|^{2}$}}}
\put(3226,-3361){\makebox(0,0)[b]{\smash{\SetFigFont{12}{14.4}{\familydefault}{\mddefault}{\updefault}$n$}}}
\end{picture}
\caption{Distributions in energy level for $\alpha=1$ ($\times$)
and $\alpha=1/32$ ($+$).}
\label{dist}
\end{center}
\end{figure}

Figure \ref{auto} depicts the autocorrelation function for the
same state as above, exhibiting the typical pattern characterizing
revivals (compare with Figure 2 of Parker and Stroud \cite{parker86}).
Figure \ref{dist} demonstrates how a small $\alpha$ shrinks the
width of the state in $n$.

Commenting again on the assertion by some authors that ``temporally
stable'' states for the hydrogen atom can not exhibit this phenomenon
\cite{bellomo98,bellomo99}.  Their calculations involved, in terms
of the present paper, $\alpha = 1$, hence wide distributions in
$n$ which exhibited no appreciable revivals over the time frames
calculated.  Wave function revivals are a universal phenomenon
depending on the nature of the energy eigenlevel spacings, and in
the case of the hydrogen atom, the width $\Delta n$.  By exploiting
the fact that one may choose $\rho(u)$ to one's liking, $\Delta n$
may be reduced such that the resultant states do revive.  Further,
one study \cite{bellomo98} used values of $s$ leading to $\langle
n \rangle$ equal to 25 and 400.  With $\langle n \rangle = 400 $,
from Eq.\ (\ref{tr}), $T_{r} \sim 5.36\times 10^{10}$ though their
calculation only extends to $t=5\times 10^{9}$.  For the other
study \cite{bellomo99}, values of $\langle n \rangle$ exhibited
were 10 and 200.  A full revival should be found using $\langle n
\rangle = 200$ at about $t=3.35 \times 10^{9}$ though their
calculation only extends to $t=5 \times 10^{8}$.

\section{Conclusion}
\label{concl}

This approach to defining temporally stable coherent states for
the hydrogen atom problem is similar in spirit to those of Majumdar
and Sharatchandra \cite{majumdar97}, Klauder \cite{klauder96}, and
Fox \cite{fox99}.  Majumdar and Sharatchandra make explicit use of
the SO(4) degeneracy group, but treat the factor $d_{n}$ differently.
Although Klauder's construction is specific to the hydrogen atom
problem rather than the present general approach, a similar term
appears in that construction which disrupts normalization.  Fox
uses a somewhat different construction which avoids the use of the
moments of some function $\rho(u)$.  Even so, one may construct
coherent states in degenerate systems related to Fox's construction
as the present construction is related to Klauder's original work.

The general construction of Eq.\ (\ref{gencs}) provides states with
many useful properties.  They form a complete set of states (in
the bound portion of the spectrum) and evolve in time among
themselves.  This makes them a clear candidate for use in
representations of time evolved, bound states.  Further, there is
a freedom in their definition which stems from the choice of
$\rho(u)$.  If $\rho(u)$ is appropriately chosen, hydrogen atom
coherent states may be defined which exhibit the full range of
phenomena exhibited by other approaches:  initial semi-classical
behaviour, interference between the head and tail of the state as it
disperses about the Keplerian ellipse, localization on the Keplerian
ellipse and wave function revivals at predictable times.

The salient difference between these coherent states and other
constructions is the is the natural and explicit manner in which
the energy degeneracies are treated herein, via Perelomov's group
theoretical construct of generalized coherent states.

\acknowledgments

The author is grateful for the support provided by the Natural Sciences
Engineering Research Council of Canada.  The author also wishes to
acknowledge useful discussions with Dr. E.R. Vrscay and Dr. J. Paldus,
both of the Department of Applied Mathematics, University of Waterloo.


\end{document}